\documentclass[aps,amsmath,amssymb,superscriptaddress,nofootinbib,preprint]{revtex4}
\usepackage{graphicx,bm}
\usepackage{axodraw}
\newcommand{\ba}{\begin{eqnarray}} \newcommand{\ea}{\end{eqnarray}}
\newcommand{\be}{\begin{equation}} \newcommand{\ee}{\end{equation}}
\newcommand{\bw}{\begin{widetext}} \newcommand{\ew}{\end{widetext}}
\newcommand{\nn}{\nonumber}

 \renewcommand{\bf}{\bfseries}

\def\21{$SU(2) \otimes U(1) $}

\def\lsim{\raise0.3ex\hbox{$\;<$\kern-0.75em\raise-1.1ex\hbox{$\sim\;$}}}
\def\gsim{\raise0.3ex\hbox{$\;>$\kern-0.75em\raise-1.1ex\hbox{$\sim\;$}}} 

\newcommand{\mx}{\left[\begin{array}}
\newcommand{\finmx}{\end{array}\right]} 
\newcommand{\mxp}{\left(\begin{array}} 
\newcommand{\finmxp}{\end{array}\right)} 
\def\beq{\begin{equation}}
\def\eeq{\end{equation}}
\def\bea{\begin{eqnarray}}
\def\eea{\end{eqnarray}}
\def\nn{\nonumber}

\def\mathbf#1{\hbox{\bf #1}}
\def\textrm#1{\hbox{#1}}

\def\lsim{\raise0.3ex\hbox{$\;<$\kern-0.75em\raise-1.1ex\hbox{$\sim\;$}}}
\def\gsim{\raise0.3ex\hbox{$\;>$\kern-0.75em\raise-1.1ex\hbox{$\sim\;$}}}

\begin{document}

\title{Leptonic color models from $Z_8$ orbifolded AdS/CFT}  

\author{K.S. Babu} 
\email{babu@okstate.edu}
\affiliation{
Department of Physics, Oklahoma State University, Stillwater,
OK 74078, USA}

\author{Thomas W. Kephart} 
\email{thomas.w.kephart@vanderbilt.edu}
\affiliation{Department of Physics and Astronomy, Vanderbilt
University, Nashville, TN 37235, USA}

\author{Heinrich P\"as} \email{hpaes@bama.ua.edu} 
\affiliation{ Department of Physics \& Astronomy, 
University of Alabama, Tuscaloosa, AL 35487, USA}

\begin{abstract}
We study orbifold compactifications of the type $IIB$
superstring on $AdS_{5}\times S^{5}/\Gamma $, where $\Gamma $ is the 
abelian  
group $Z_8$, which can lead to non-SUSY three and four family models 
based on quartification. In particular, we focus on two models, one 
fully quartified model and one a model with two trinification 
families and one quartification family, which reduces to 
the standard model with a minimal leptonic color sector. 
\end{abstract}

\maketitle

\date{today}

\section{Introduction}
Orbifold compactifications of the type $IIB$
superstring on $AdS_{5}\times S^{5}$ \cite{Maldacena:1997re,Lawrence:1998ja}
(for a review see \cite{Frampton:2007fr}) 
lead to gauge theories with $SU^n(N)$
gauge groups
when the orbifolding group is abelian and of order $n$. Trinification models
\cite{deRujula,Babu:1985gi} with gauge group $SU(3)_L \times SU(3)_C \times
SU(3)_R$ 
and 
quartification  
\cite{JV,Babu:2003nw,Chen:2004jz,Demaria:2005gk,Demaria:2006uu,Demaria:2006bd} 
models, where 
the gauge group is extended to 
$SU(3)_l \times SU(3)_L \times SU(3)_C \times SU(3)_R$ 
are of this class. Thus a natural question to ask is 
whether one can derive models with the appropriate fermion content to allow 
for three families of quarks and leptons, and the appropriate scalar content 
to permit gauge symmetry breaking to the standard model and ultimately to 
$SU(3)_C\times U_{EM}(1)$. In \cite{Kephart:2001qu,Kephart:2004qp} two of us 
carried out a global search for  $\Gamma=Z_n $ trinification models with 
three or 
more families. Here we will concentrate on phenomenologically interesting 
quartification models. These models contain a leptonic color sector 
to realize a manifest quark-lepton symmetry 
\cite{Foot:dw,Foot:fk,Foot:2006ie}
and must contain at 
least three normal families to be phenomenologically viable, plus they 
contain the new fermions needed to symmetrize the quark and lepton particle 
content at high energies. We will consider both models 
with the full quartification (all families are quartification families),
\be{}
3 (3\bar{3}11)+ (13\bar{3}1) +  (113\bar{3}) +  (\bar{3}113)
\ee
 and 
hybrid models where two families are trinification families and the third is 
a quartification family,
\be{}
2 [(3\bar{3}1) + (13\bar{3}) + (\bar{3}13)] +
(3\bar{3}11)+ (13\bar{3}1) +  (113\bar{3}) +  (\bar{3}113).
\ee
(Potentially, the family splitting could also be 
one trinification 
plus two quartification families.)
The plan of the paper is as follows. We first review generic 
trinification and 
quartification models. Next we review the rules for generating 
models based on 
orbifold compactifications of the type $IIB$ superstring. We then restrict 
ourselves to  $\Gamma=Z_8 $ where we find the first phenomenologically 
interesting quartification  and trinification--quartification hybrid models. 
We next study a specific semi-realistic four family quartification model and 
then 
an even more promising trinification--quartification hybrid model. We end 
with a discussion and summary of our main results.

\section{Review of trinification and quartification models}
Trinification models  \cite{deRujula}  are based on the 
gauge group $SU(3)_C \times SU(3)_L \times SU(3)_R$ where the electric 
charge operator is
\begin{equation}
Q = I_{3L} - {Y_L \over 2} + I_{3R} - {Y_R \over 2} = I_{3L} + {Y \over 2}.
\end{equation}
In terms of $SU(3)_L \times  SU(3)_R$, the leptons are 
\begin{equation}
 \ell \sim 
 \begin{pmatrix} 
 N & E^c & \nu \cr E & N^c & e \cr \nu^c & e^c & S,
  \end{pmatrix},
\end{equation}
where $I_{3L} = (1/2, -1/2, 0)$ and $Y_L = (1/3, 1/3, -2/3)$ for the rows,
and  $I_{3R} = (-1/2, 1/2, 0)$ and $Y_R = (-1/3, -1/3, 2/3)$ for the columns
and the
exotic fermion $h(h^c)$,  $E(E^c)$, and $N,N^c,S$ have charges
$\mp 1/3$, $\mp 1$, and  0 respectively. 
The quarks can also be arranged in matrix form
\begin{equation}
q \sim 
 \begin{pmatrix} 
 d & u & h \cr d & u & h \cr d & u & h
   \end{pmatrix},
  ~~~ q^c \sim 
 \begin{pmatrix}
d^c & d^c & d^c \cr u^c & u^c & u^c \cr h^c & h^c & h^c
  \end{pmatrix},
\end{equation}
where $I_{3L} = (-1/2, 1/2, 0)$, 
$Y_L =(-1/3, -1/3, 2/3)$
for the columns in $q$ and 
$I_{3R} = (1/2,-1/2, 0)$,
 $Y_L =(1/3, 1/3, -2/3)$
for the rows in $q^c$. 
More compactly
\begin{eqnarray}
&~ l \sim (1,3,\bar{3}),\\
q \sim (3,\bar{3},1), &~& q^c \sim (\bar{3},1,3).
\end{eqnarray}
Vacuum expectation values (VEVs) for two scalar multiplets
$\phi_a \sim (1,3,\bar{3})~~(a=1,2)$
provide appropriate fermion masses and mixings. 
Trinification models can be nicely unified into $E_6$ but are not symmetric
between their quark and lepton content. 

Quartification models 
\cite{{JV}, {Babu:2003nw}}, 
where the gauge 
group is extended to $SU(3)_l \times SU(3)_L \times SU(3)_C \times SU(3)_R$, 
have quark-lepton symmetry, where 
\begin{eqnarray}
l \sim (3,\bar{3},1,1), &~& l^c \sim (\bar{3},1,1,3),\\
q \sim (1,3,\bar{3},1), &~& q^c \sim (1,1,3,\bar{3}).
\end{eqnarray}
The electric charge operator 
becomes
\begin{equation}
Q = I_{3L} - {Y_L \over 2} + I_{3R} - {Y_R \over 2} - {Y_l \over 2}.
\end{equation}
Here $Y_l$ takes the same values as $Y_L$ or $Y_R$ depending of course on
whether it is part of a triplet or antitriplet.  The matrix representations 
of
$l$ and $l^c$ are
\begin{equation}
l \sim \begin{pmatrix} x_1 & x_2 & \nu \cr y_1 & y_2 & e \cr z_1 & z_2 & N
\end{pmatrix}, ~~~ l^c
\sim \begin{pmatrix} x_1^c & y_1^c & z_1^c \cr x_2^c & y_2^c & z_2^c \cr 
\nu^c & e^c & N^c
\end{pmatrix}.
\end{equation}
Here
the columns of $l$ have $Y_l = (-1/3,-1/3,2/3)$ and the rows have
$I_{3L} = (1/2, -1/2, 0)$, 
$Y_L =(1/3, 1/3, -2/3)$. 
The rows of $l^c$
have $Y_l$ = (1/3,1/3, -2/3), and the columns have
$I_{3R} = (-1/2,1/2, 0)$,
$Y_L =(-1/3, -1/3, 2/3)$.
 
Of the new particles $N$ and $N^c$ are
neutral. The exotic $SU(2)_l$ doublet leptons $(x,y,z)$ have charges
$(1/2,-1/2,1/2)$ and $(x^c,y^c,z^c)$ have charges $(-1/2,1/2,-1/2)$,
respectively  \cite{Babu:2003nw}.  
Because of their half integral charges, the $SU(2)_l$
doublets have been dubbed ``hemions''.

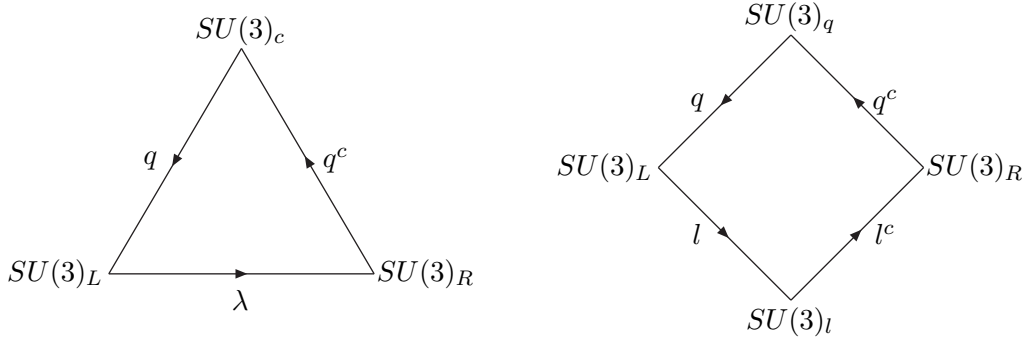
\begin{figure}[!t]
\begin{center}
\parbox{7cm}{
\begin{picture}(300,100)(0,0)
\ArrowLine(65,10)(165,10) \ArrowLine(115,95)(65,10) \ArrowLine(165,10)(115,95)
\Text(115,102)[]{$SU(3)_c$} \Text(45,10)[]{$SU(3)_L$}
\Text(185,10)[]{$SU(3)_R$} \Text(81,53)[]{$q$} \Text(151,53)[]{$q^c$}
\Text(115,0)[]{$\lambda$}
\end{picture}}
\parbox{7cm}{
\begin{picture}(300,110)(0,0)
\ArrowLine(120,5)(170,55) \ArrowLine(170,55)(120,105) \ArrowLine(70,55)(120,5)
\ArrowLine(120,105)(70,55) \Text(120,112)[]{$SU(3)_q$}
\Text(120,-3)[]{$SU(3)_l$} \Text(50,55)[]{$SU(3)_L$} \Text(190,55)[]{$SU(3)_R$}
\Text(85,80)[]{$q$} \Text(156,80)[]{$q^c$} \Text(85,30)[]{$l$}
\Text(156,30)[]{$l^c$}
\end{picture}}
\end{center}
\caption{Quiver diagrams of $[SU(3)]^3$ trinification and $[SU(3)]^4$ 
quartification.}
\label{quivers}
\end{figure}

Symmetry breaking of quartification models to the standard model 
can be somewhat involved so we will 
delay a discussion until we get to specific models derived from AdS/CFT, but 
we note here that the fermions all fall into bifundamental representations of 
the gauge group, and these are naturally arranged into a moose or quiver
diagram
\cite{Georgi:1985hf,Douglas:1996sw}, see Fig.~\ref{quivers}.

\section{Rules for $AdS_{5}\times S^{5}/Z_n$ model building}

In this paper we study conformal field theory models originating from
the large $N$ expansion of the AdS/CFT \cite{Maldacena:1997re} correspondence.
We choose $N=3$ and as a consequence, the gauge group of the corresponding
CFT model derived from an abelian orbifold
is given by the product group $SU^n(3)$.
The ${\cal N} = 4$ supersymmetry of $AdS_{5}\times S^{5}$ is broken upon 
orbifolding $S^5 \rightarrow S^5/\Gamma$ where $\Gamma$ is a finite group
embedded in the isometry
$SU(4) \sim O(6)$ of $S^5$. Here we concentrate on the case $\Gamma=Z_n$
with $n=8$. 
The choice of $Z_8$ is not arbitrary, in fact a systematic search of $Z_n$
orbifold models reveals that $Z_8$ is the minimal choice on which to base a
phenomenological quartification model in which all product groups have the 
same coupling strength.
The {\bf 4} of the $SU(4)$ isometry
must be neither real nor pseudoreal for chiral fermions to be
present in the resulting quiver gauge theory. The number of surviving
supersymmetries is ${\cal N} = 2, 1, 0$ for 
$\Gamma$ embedded nontrivially in $SU(2)$, $SU(3)$, or $SU(4)$ respectively.
For $\alpha = \exp(2 \pi i / n)$ the
embedding is fixed by a choice 
{\rm{\bf 4}} $=(\alpha^{A_1}, \alpha^{A_2}, \alpha^{A_3},
\alpha^{A_4})$. We denote such a model as
$M_{A_1A_2A_3A_4}$ and define it as partition or double partition model, if
$A_1 + A_2 + A_3 +A_4=n$ or $2n$, respectively
 (see 
\cite{Kephart:2001qu,Kephart:2004qp} for notational details).
Partition or double partition models are particular attractive, as
the construction of viable string theory 
non-partition models may not be possible \cite{Frampton:2003vc}.  
For ${\cal N} = 0$, the case of interest here,  all four $A_i$ 
need to be nontrivial. This determines the fermion content of the theory:
fermions reside in the bilinear representations of the 1st and $A_i$-th
product group and its cyclic permutations.
Consider next the {\bf 6} of $SU(4)$ which is the antisymmetric
part in {\bf 4}$\times${\bf 4}.
Consistency \cite{Frampton:2003vc} requires a real embedding of the 
{\bf 6} which can be written in the form 
{\bf 6} $=(A_1+A_4, A_2+A_4, A_3+A_4, A_1+A_2, A_2+A_3, A_3+A_1)$. 
This in turn determines the scalar content of the theory.

\section{Leptonic color models from AdS/CFT}

It has been shown \cite{Kephart:2004qp}
that the viable $Z_8$ orbifolds of AdS/CFT include
five partition models and one double partition model.
In this work we study the symmetry breaking to the quartification group
\be
SU(3)_l \times SU(3)_L \times SU(3)_C \times SU(3)_R,
\ee
where each factor is the diagonal subgroup of two of the original
$SU(3)^8$ factor groups. This procedure yields quartification models in
which the individual factor groups initially
couple with the same strengths.

In general there exist 24 different symmetry breaking patterns of this type.
A systematic search yielded only two viable models.
One, a $M_{1133}$ model, is a semi-realistic, four family quartification 
model and the other is an 
$M_{1456}$ model, which leads to a more phenomenologically interesting
hybrid fermion spectrum with two trinification families and one 
quartification family. 

\subsection{Quark-lepton quartification from AdS/CFT}

The $SU(3)^8$ fermion spectrum of $M_{1133}$ is given by
\be{}
2 [(3\bar{3}111111)+ (311\bar{3}1111)]_F + {\rm cyclic~~permutations} 
\ee
and the scalar spectrum is given by
\be{}
 [(31\bar{3}11111) + 4 (3111\bar{3}111) + (311111\bar{3}1) + h.c.]
+ {\rm cyclic~~permutations}.
\ee
We can break $SU(3)^8$ to $SU(3)^4$ 
by assigning VEVs to scalars in the representations
$(31\bar{3}11111)$, $(131\bar{3}1111)$, $(111131\bar{3}1)$ and
$(1111131\bar{3})$. Omitting vectorlike fermions, 
singlets and octets the chiral $SU(3)^4$ fermions are
\be{}
4[(3\bar{3}11)+(13\bar{3}1)+(113\bar{3})+(\bar{3}113)]_F
\ee 
and the scalars are 
\be{}
10[(31\bar{3}1)+(131\bar{3})+h.c.]_S+2 [(8111)+(1811)+(1181)+(1118)]_S.
\ee
Note that the chiral fermion content of this model is precisely four 
quartification families. 

To proceed toward the standard model we first label the four remaining 
$SU(3)$'s as 
$SU(3)_l \times SU(3)_L \times SU(3)_C \times SU(3)_R$. With an octet $(1811)$ 
who's VEV is proportional to $\lambda_8$ we can break $SU(3)_L$ to 
$SU(2)_L\times U_L(1)$. A second $(1811)$ with VEV proportional to 
$\lambda_1$ allows us to break $SU(2)_L$ completely but leave $U_L(1)$ 
unbroken. 
Likewise, VEVs for two octets of type $(1118)$ allows us to break $SU(3)_R$ 
down to $U_R(1)$. Next a $(131\bar{3}) + h.c.$ can be used to break 
$U_L(1)\times U_R(1)$ to the diagonal subgroup $U_D(1)$. To achieve the final
symmetry of the quartification we still need to break  $SU(3)_l $. 
First a 
$\lambda_8$ type octet $(8111)$ VEV gives $SU(2)_l\times U_l(1)$. 
The $SU(2)_l$ needs to remain unbroken, but $U_D(1)\times U_l(1)$ is 
required to break to a linear combination that is the weak hypercharge. 
There are no remaining scalar representations that can do this, but it is 
possible that a leptonic color condensate forms that has both $U_D(1)$ and 
$U_l(1)$ charge, reducing the symmetry to the desired linear combination. 
However, since the $SU(2)_l$ scale 
$\Lambda_{lCD}$ is much below the color confinement scale determined by 
$\Lambda_{QCD}$, the formation of such a condensate provides a proof of
principle rather that a viable phenomenology. We must proceed to the the 
$M_{1456}$ model if that is what we desire.

\subsection{Minimal leptonic color from AdS/CFT} 

We now consider  the more realistic double partition model $M_{1456}$:
the particle spetrum of the unbroken $SU(3)^8$ theory at the string scale
is given by
\be{}
(3\bar{3}111111)_F+(3111\bar{3}111)_F+(31111\bar{3}11)_F+(311111\bar{3}1)_F
h.c. + {\rm cyclic~~permutations}
\ee 
fermion states and scalars in the
\begin{eqnarray}
&(3\bar{3}111111)_S+(31\bar{3}11111)_S+(311\bar{3}1111)_S+
(31111\bar{3}11)_S+(311111\bar{3}1)_S+(3111111\bar{3})_S & \nonumber\\
&+{\rm cyclic~~permutations}&
\end{eqnarray}
representations.
$SU(3)^8$ is broken down to $SU(3)^4$ by assigning VEVs to   
$(3\bar{3}111111)_S$,
$(11311\bar{3}11)_S$,
$(111311\bar{3}1)_S$ and
$(1111311\bar{3})_S$ which leaves 
chiral fermions now in the representations
\be{}
2[(13\bar{3}1) + (113\bar{3}) + (1\bar{3}13)]_F 
+ [(3\bar{3}11)+(13\bar{3}1) + (113\bar{3}) + (\bar{3}113)]_F,
\ee
and scalars in the representations
\ba{}
3[(3\bar{3}11)+(13\bar{3}1) + (113\bar{3}) + (\bar{3}113) + h.c.]_S 
+4[(31\bar{3}1)+(131\bar{3}) + h.c.]_S   \nn \\
+2[(8111)+(1811)+(1181)+(1118)]_S.  
\label{scalars}
\ea

We assume the two light families are  the trinification families and the 
heavy
quarks are in the color (anti-)triplet quartification multiplets.
The third family leptons plus exotic matter can be obtained from the
leptonic color (anti-)triplets.

We could give a VEV to a $(3\bar{3}11)$ and immediately break 
$SU(3)_l \times 
SU(3)_L$ to a new $SU(3)_L'$ which would yield a three family trinification 
model. We choose not to do this, as these models have already been 
extensively 
explored. Instead, in analogy with \cite{Babu:2003nw}, and
with the labeling 
$SU(3)_l \times SU(3)_L \times SU(3)_C \times SU(3)_R$, 
VEVs are given to two $(131\bar{3})$ representations
to generate realistic quark masses
and non-zero mixing angles, and to $(3\bar{3}11)$ and to 
$(311\bar{3})$, by which 
$SU(3)_l$ is broken down to $SU(2)_l$.
This way the scalar content given in (\ref{scalars}) results in three Higgs
doublets from the $(3\bar{3}11)$ representations and another 12 Higgs doublets 
from the $(131\bar{3})$ representations, which will be important for the
phenomenology of the model.

Alternatively,
the octets of type 
$(1811)_S$ and $(1118)_S$ are again sufficient to break 
$SU(3)_L \times SU(3)_R$ to $SU(2)_L \times U(1)_Y$. At this stage $SU(3)_l$ 
remains unbroken. It is interesting to note that at the $SU(3)_l \times
SU(3)_L \times SU(3)_C \times SU(3)_R$ level, the $\beta$ function for
$SU(3)_l$ 
is the most negative. Ignoring the 
scalars for the moment, the fermionic 
term in $\beta_l$ is $4/3 N_l$ where $N_l=3$ is the number of $SU(3)_l $ 
triplet Dirac fermions, while for $ SU(3)_L \times SU(3)_C \times SU(3)_R$ 
we find fermionic terms with $N_L=9$, $N_C=9$, and $N_R=9$. 
Consequently, $SU(3)_l$ is asymptoticly free and its coupling constant 
$\alpha_l$ becomes of order one far above $\Lambda_{QCD}$.

\section{Phenomenological Consequences}

\begin{figure}[t]
\centering
\includegraphics[clip,scale=0.8]{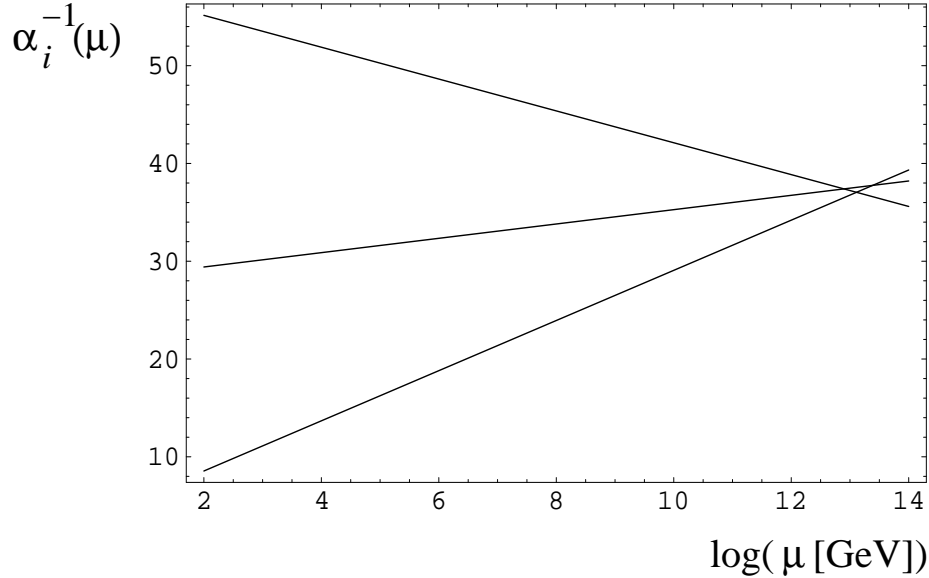}
     \caption{Gauge coupling unification in the minimal leptonic color model,
              assuming two trinification and one quartification families of
              fermions, and 4 Higgs doublets. The couplings unify around  
              $10^{13}$~GeV.
     \label{couplings}}
\end{figure}

In the following we sketch some phenomenological aspects of the minimal
leptonic color model. We follow here the discussion of the original
quartification model in \cite{Babu:2003nw}, but stress several interesting
new aspects due to the two incomplete quartification families and
the remnant particles from the $SU(3)^8 \rightarrow SU(3)^4$ symmetry
breaking,

{\it Gauge coupling unification:}
The renormalization-group evolution of the gauge couplings in leading order
is given by
\begin{equation}
\frac{1}{\alpha_i(\mu)}-\frac{1}{\alpha_i(\mu')}=\frac{b_i}{2\pi}
\ln\left(\frac{\mu'}{\mu}\right)\;,\label{rng}
\end{equation}
where $b_n$ are the one-loop beta-function coefficients,
\begin{eqnarray}
&&b_3 = -11 + \frac{4}{3}N_g, \\
&&b_2 = -\frac{22}{3} +2 N_q + \frac{4}{3} N_t + \frac{1}{6}N_H, \\
&&b_1 = \frac{13}{9}N_q + \frac{4}{3}N_t + \frac{1}{12}N_H.
\label{renorm}
\end{eqnarray}
Here $N_g=3$ is the number of generations and $N_q=1$, $N_t=2$ accounts for the
number of quartification and trinification families.
The running includes the contributions of the exotic
weak-scale $SU(2)_l$ doublet ``hemions'' [$(x,y)$ is an $SU(2)_L$ doublet with
$Y=0$; $x^c$ and $y^c$ are $SU(2)_L$ singlets with $Y=\mp1$] and $N_H$ Higgs
doublets with $Y=\pm 1$.  The initial values of the gauge couplings are
\begin{eqnarray}
&&\alpha_3(M_Z)=0.117, \\
&&\alpha_2(M_Z)=(\sqrt 2/\pi)G_FM_W^2 = 0.034, \\
&&\alpha_1(M_Z) = \alpha_2(M_Z)
\left(\tan^2\theta_W/\tan^2\theta_W(M_{GUT})\right) = 0.0181,
\end{eqnarray}
where $\sin^2\theta_W(M_{GUT})=\sum I^2_{3L}/\sum Q^2 = 9/16$
(the sum running over all fermion representations) 
is determined by the embedding of $U(1)_Y$ in $[SU(3)]^4$. 
There are a total of 12 standard (uncolored)
doublets of $SU(2)_L$. There are also 9 doublets with color and 9 more
with leptocolor. All the standard doublets will be able to grow a
mass when we have broken to the SM gauge group (or to SM gauge group $\times$
leptocolor), but we could keep them light by fine tuning, or by
keeping the $U(1)$'s from $SU(3)_L$ and $SU(3)_R$ unbroken. This is similar 
to
the symmetry breaking
$E_6 \rightarrow SU(3)^3 \rightarrow SU(3)\times
SU(2) \times U(1) \times U(1)' \times U(1)''$. 
As long as we keep all three 
$U(1)$s
unbroken, all components of the 27 of $E_6$ remain massless.
The evolution of the couplings from the weak scale up to very high scales
is shown
in Fig.~\ref{couplings}, using $N_H=4$. The gauge couplings unify around  
$10^{13}$~GeV, at a somewhat higher energy scale as compared to
\cite{Babu:2003nw}. An interesting consequence would be the chance to discover
multiple Higgs doublets at the LHC.

For the coupling of the unbroken $SU(2)_l$ group the one-loop beta-function 
coefficient is given by
\be{}
b_{2l}=-\frac{22}{3}+\frac{4}{3}N_q.
\ee
The trinification families don't contribute as they are singlets under
$SU(2)_l$. This implies $\alpha^{-1}_{2l}(M_Z)\simeq 13$, which is between the
weak and the strong couplings and will yield a similar phenomenology as in
\cite{Babu:2003nw}, albeit with a higher scale, where the leptonic color
interaction becomes non-perturbative, somewhat below an MeV.

\noindent
{\it Hemion masses:}
As in \cite{Babu:2003nw}, quartification scale hemion masses are forbidden
by the $Z_8$ orbifold symmetry. 
TeV scale hemion masses could be generated by adding non-renormalizable
operators that are suppressed by the Planck scale.

\noindent
{\it Electroweak precision data:}
Generally one should worry about electroweak precision data in view of the
variety of new particles introduced by the model. However, the singlet and
vectorlike symmetry breaking products will not affect these processes, and the 
hemions are vector-like under the SM gauge group, thereby their contribution
would be suppressed by the hemion masses. The same argument applies to the
Higgs sector. An important issue could be the discussion of potentially 
excessive flavor changing neutral currents, but this is beyond the scope of 
this work and will be studied elsewhere.

\noindent
{\it Proton decay:}
As usual in product group unifications scenarios,
proton decay will not be mediated by gauge bosons.
However, proton decay could be induced via couplings to the extended scalar
sector given in (\ref{scalars}), see \cite{Dent:2007ed}.

\noindent
{\it Neutrino masses:}
The symmetry breaking chain discussed in section IV B provides several 
singlet fermions, e.g. from the $(3\bar{3}111111)$ representation after
breaking the first two $SU(3)$ 
groups down to the diagonal subgroup by assigning
a VEV to the scalar representation $(3\bar{3}111111)$. This makes the string
inspired model superior to the simple quartification model, as 
a seesaw mechanism
\be{}
m_\nu \sim \frac{m^2_{\bar{\nu}\nu}}{M_{3\bar{3}111111}}
\ee
can be implemented to generate light neutrino masses without adding
right-handed
neutrinos by hand to the theory.

\noindent
{\it Stickballs}
are the glueballs of the leptonic color SU(2). These particles
could act as a cold dark matter candidate.

\section{Discussion}

We have shown that it is possible to find quartification models based on 
orbifold compactifications of the type $IIB$ superstring on 
$AdS_{5}\times S^{5}/Z_n $. These models have fermions in only bifundamential
representations and can have a sufficient number of scalar fields to allow 
spontaneous symmetry breaking to the standard model. The first two models of 
this type arise at $n=8$. The first fully quartified model is somewhat 
less than realistic since the final stage of symmetry breaking relies on 
condensates to provide a proof of principle rather than a viable 
phenomenology. The other model does have sufficient number of scalar 
fields to allow the complete spontaneous symmetry breaking to the standard 
model. It has the interesting additional feature that one family is fully 
quartified, while the other two families are of the trinification type. 
This suggests the possibility of a natural family hierarchy in this hybrid 
model
which could lead to interesting phenomenology, including a rich particle
spectrum within reach of the LHC.

\section*{Acknowledgments}
TWK was supported by the US Department of Energy under Grant
DE-FG05-85ER40226. HP was supported
by the US Department of Energy under Grant DE-FG03-91ER40833, by the
the University of Alabama and the EU project ILIAS N6 WP1. 
TWK and HP thank the Aspen Center for Physics for hospitality while
this research was in progress.


\begin{thebibliography}{99}

\bibitem{Maldacena:1997re}
  J.~M.~Maldacena,
  %
  Adv.\ Theor.\ Math.\ Phys.\  {\bf 2}, 231 (1998)
  [Int.\ J.\ Theor.\ Phys.\  {\bf 38}, 1113 (1999)]
  [arXiv:hep-th/9711200].

\bibitem{Lawrence:1998ja}
  A.~E.~Lawrence, N.~Nekrasov and C.~Vafa,
  Nucl.\ Phys.\  B {\bf 533}, 199 (1998)
  [arXiv:hep-th/9803015].
  
\bibitem{Frampton:2007fr}
  P.~H.~Frampton and T.~W.~Kephart,
  arXiv:0706.4259 [hep-ph].

\bibitem{deRujula}
S.~L.~Glashow,
Print-84-0577 (BOSTON);
A.~De R\'ujula, H.~Georgi, and S.~L.~Glashow, in {\sl Fifth Workshop on Grand
Unification}, ed.~K.~Kang, H.~Fried, and P.~Frampton (World Scientific,
Singapore, 1984), p.~88.

\bibitem{Babu:1985gi}
K.~S.~Babu, X.~G.~He and S.~Pakvasa,
Phys.\ Rev.\ D {\bf 33}, 763 (1986);
X.~G.~He and S.~Pakvasa,
Phys.\ Lett.\ B {\bf 173}, 159 (1986);
H.~Nishimura and A.~Okunishi,
Phys.\ Lett.\ B {\bf 209}, 307 (1988);
E.~D.~Carlson and M.~Y.~Wang,
arXiv:hep-ph/9211279;
G.~Lazarides, C.~Panagiotakopoulos, and Q.~Shafi,
Phys.\ Lett.\ B {\bf 315}, 325 (1993) [Erratum-ibid.\ B {\bf 317}, 661 
(1993)]
[arXiv:hep-ph/9306332].
G.~Lazarides and C.~Panagiotakopoulos,
Phys.\ Lett.\ B {\bf 336}, 190 (1994)
[arXiv:hep-ph/9403317];
G.~Lazarides and C.~Panagiotakopoulos,
Phys.\ Rev.\ D {\bf 51}, 2486 (1995)
[arXiv:hep-ph/9407286];
S.~Willenbrock,
Phys.\ Lett.\ B {\bf 561}, 130 (2003)
[arXiv:hep-ph/0302168];
K.~S.~Choi and J.~E.~Kim,
Phys.\ Lett.\ B {\bf 567}, 87 (2003)
[arXiv:hep-ph/0305002];
J.~E.~Kim,
Phys.\ Lett.\ B {\bf 591}, 119 (2004)
[arXiv:hep-ph/0403196];
C.~D.~Carone and J.~M.~Conroy,
Phys.\ Rev.\ D {\bf 70}, 075013 (2004)
[arXiv:hep-ph/0407116];
C.~D.~Carone,
Phys.\ Rev.\ D {\bf 71}, 075013 (2005)
[arXiv:hep-ph/0503069];
A.~Demaria and R.~R.~Volkas,
Phys.\ Rev.\ D {\bf 71}, 105011 (2005)
[arXiv:hep-ph/0503224];
C.~D.~Carone and J.~M.~Conroy,
Phys.\ Lett.\ B {\bf 626}, 195 (2005)
[arXiv:hep-ph/0507292].
 

\bibitem{JV}
G.~C.~Joshi and R.~R.~Volkas,
Phys.\ Rev.\ D {\bf 45}, 1711 (1992).

\bibitem{Babu:2003nw}
K.~S.~Babu, E.~Ma and S.~Willenbrock,
{\it Quark lepton quartification},
Phys.\ Rev.\ D {\bf 69}, 051301 (2004)
[arXiv:hep-ph/0307380].

\bibitem{Chen:2004jz}
  S.~L.~Chen and E.~Ma,
  Mod.\ Phys.\ Lett.\  A {\bf 19}, 1267 (2004)
  [arXiv:hep-ph/0403105].

\bibitem{Demaria:2005gk}
  A.~Demaria, C.~I.~Low and R.~R.~Volkas,
  Phys.\ Rev.\  D {\bf 72}, 075007 (2005)
  [Erratum-ibid.\  D {\bf 73}, 079902 (2006)]
  [arXiv:hep-ph/0508160].

\bibitem{Demaria:2006uu}
  A.~Demaria, C.~I.~Low and R.~R.~Volkas,
  Phys.\ Rev.\  D {\bf 74}, 033005 (2006)
  [arXiv:hep-ph/0603152].

\bibitem{Demaria:2006bd}
  A.~Demaria and K.~L.~McDonald,
  Phys.\ Rev.\  D {\bf 75}, 056006 (2007)
  [arXiv:hep-ph/0610346].

\bibitem{Kephart:2001qu}
T.~W.~Kephart and H.~P\"as,
Phys.\ Rev.\ D {\bf 70}, 086009 (2004)
[arXiv:hep-ph/0402228].

\bibitem{Kephart:2004qp}
T.~W.~Kephart and H.~P\"as,
Phys.\ Lett.\ B {\bf 522}, 315 (2001)
[arXiv:hep-ph/0109111];

\bibitem{Foot:dw}
R.~Foot and H.~Lew,
Phys.\ Rev.\ D {\bf 41}, 3502 (1990).

\bibitem{Foot:fk}
R.~Foot, H.~Lew, and R.~R.~Volkas,
Phys.\ Rev.\ D {\bf 44}, 1531 (1991).

\bibitem{Foot:2006ie}
  R.~Foot and R.~R.~Volkas,
  arXiv:hep-ph/0607047.

\bibitem{Georgi:1985hf}
H.~Georgi,
Nucl.\ Phys.\ B {\bf 266}, 274 (1986).

\bibitem{Douglas:1996sw}
M.~R.~Douglas and G.~W.~Moore,
arXiv:hep-th/9603167.


\bibitem{Frampton:2003vc}
  P.~H.~Frampton and T.~W.~Kephart,
  Int.\ J.\ Mod.\ Phys.\ A {\bf 19}, 593 (2004)
  [arXiv:hep-th/0306207].


\bibitem{footnote}
State counting here gives $[6,18(+6),18,18(+6)]$ compared to
$[18,18,18,18]$ above.

\bibitem{Dent:2007ed}
  J.~B.~Dent and T.~W.~Kephart,
  arXiv:0704.1451 [hep-ph].

\end{thebibliography}
\end{document}